\documentclass[12pt]{article}
\usepackage{graphicx,epsfig,color}
\usepackage{cite}
\renewcommand{\thefootnote}{\fnsymbol{footnote}}

\newcommand{\bc}{\begin{center}}
\newcommand{\ec}{\end{center}}
\newcommand{\be}{\begin{equation}}
\newcommand{\ee}{\end{equation}}
\newcommand{\bea}{\begin{eqnarray}}
\newcommand{\eea}{\end{eqnarray}}
\newcommand{\ba}{\begin{array}}
\newcommand{\ea}{\end{array}}
\newcommand{\lb}{\label}
\newcommand{\rf}{\ref}
\newcommand{\bfg}{\begin{figure}[htbp]}
\newcommand{\efg}{\end{figure}}

\newcommand{\pr}{Phys. Rev. }
\newcommand{\prd}{Phys. Rev. D }
\newcommand{\np}{Nucl. Phys. }
\newcommand{\npb}{Nucl. Phys. B }
\newcommand{\prl}{Phys. Rev. Lett. }
\newcommand{\prp}{Phys. Rep. }
\newcommand{\ap}{Ann. Phys. (N.Y.) }

\newcommand{\plb}{Phys. Lett. B }

\newcommand{\rmp}{Rev. Mod. Phys. }

\newcommand{\epjc}{Eur. Phys. J. C }
\newcommand{\jhep}{JHEP }

\begin{document} 

\vspace*{1. cm}
\bc
{\Large {\boldmath \textbf{Cluster reducibility of multiquark
operators}}}
\\
\vspace{1. cm}
{Wolfgang Lucha$^a$, Dmitri Melikhov$^{a,b,c}$, Hagop Sazdjian$^d$}\\
\vspace{0.5 cm}
{\small
$^a$Institute for High Energy Physics, Austrian Academy of 
Sciences, Nikolsdorfergasse 18,\\ 
A-1050 Vienna, Austria\\
$^b$D.~V.~Skobeltsyn Institute of Nuclear Physics,
M.~V.~Lomonosov Moscow State University,\\ 
119991, Moscow, Russia\\
$^c$Faculty of Physics, University of Vienna, Boltzmanngasse 5, 
A-1090 Vienna, Austria\\ 
$^d$Institut de Physique Nucl\'eaire, Universit\'e Paris-Sud,
CNRS-IN2P3,\\
Universit\'e Paris-Saclay, 91405 Orsay, France\\
E-mail:  wolfgang.lucha@oeaw.ac.at , dmitri\_melikhov@gmx.de , 
sazdjian@ipno.in2p3.fr 
}
\ec
\par
\vspace{0.25 cm}

\bc
{\large Abstract}
\ec
\par
It is shown that the multiquark gauge-invariant operators can, in
general, be decomposed into combinations of products of ordinary
hadronic operators, exhibiting their cluster reducibility.
The latter property inhibits the formation of completely compact
multiquark bound states. Multiquark operators still play a
crucial role in the description of exotic states in regions of
configuration space where the hadronic clusters are close to each
other. Our proof gives a foundation for a unified viewpoint, where
the multiquark-type and the molecular-type approaches play
complementary roles, at the gauge-invariant nonlocal operator level.
\par

\vspace{0.25 cm}
{\small
PACS numbers: {11.15.Pg, 12.38.Lg, 12.39.Mk, 13.25.Jx, 14.40.Rt} 
\par
Keywords: QCD, multiquark operators, tetraquarks, pentaquarks,
hexaquarks. 
}
\par 
\vspace{0.25 cm}
\renewcommand{\thefootnote}{\arabic{footnote}}

\section{Introduction} \lb{s1}
\setcounter{equation}{0}
\setcounter{figure}{0}

The possibility of the existence of multiquark states
\cite{Jaffe:1976ih,Jaffe:2008zz}, i.e., of states containing more than
a pair of valence quark-antiquark for mesons and more than three
valence quarks for baryons, also called exotic states, raises in QCD
the following question: can QCD produce, by the sole confining forces,
multiquark bound states in the same way as it produces ordinary
hadrons?
\par
An indication that multiquark states are not of the completely
confined type arises from the observation that interpolating local
currents, that might couple to them, can always, by Fierz
rearrangements, be brought into forms where they appear as
combinations of products of color-singlet quark bilinear and/or
trilinear local operators \cite{Jaffe:2008zz,Coleman:1985rnk}.
As a consequence, correlation functions of such currents become
dominated by disconnected diagrams, representing free hadron
propagators, while the connected diagrams depict interactions among
hadrons, which are of the nonconfining type.
\par
It might seem that the latter property concerns only some particular
aspects of multiquark states, since their couplings to local
currents involve only a few moments of their wave functions.
A wider view of their structure is provided by the general
gauge-invariant states constructed with the aid of path-ordered
gluon-field phase factors, also called Wilson lines, which are
the color parallel transporters of the theory. One thus naturally
arrives at the ``string-junction'' or ``$Y$-shaped-junction''
representation of multiquark states
\cite{Rossi:1977cy,Rossi:2016szw}, also characterized as ``compact''
states. Here, one expects to exhibit more easily their confined
nature by means of their bound-state spectrum, which should show
up through a tower of states lying at non-negligible distances
from the multiquark thresholds, in analogy with the
ordinary-hadron cases. A realization of the $Y$-shaped-type
interactions is provided by the diquark picture \cite{Jaffe:2003sg,
Shuryak:2003zi,Maiani:2004vq}.
\par
The purpose of the present article is to show that even the
general gauge-invariant multiquark operators are cluster reducible,
in the sense that they are decomposable into a combination of products
of gauge-invariant bilinear or trilinear operators, reminiscent of
ordinary hadronic operators, thus generalizing the phenomenon
encountered with local interpolating currents\footnote{The
cluster reducibility property of multiquark operators was first
emphasized in 1980 by Jan Stern, who called attention, through
seminars, to that aspect. He did not, however, leave any written
article about the subject. The proof presented in this article is
based on the arguments developed by Jan Stern.}. 
\par
The cluster reducibility property of multiquark operators does not
leave enough room for the occurrence of a possible globally confined
structure of the corresponding states, since interactions among
hadronic clusters are expected to be nonconfining.
Possible implications of this result and connections with other
investigations about multiquark states will be discussed in Sec. 4.
Sections 2 and 3 are devoted to the details of the proof of cluster
reducibility in the $SU(3)$ and $SU(N_c^{})$ cases, respectively.
A summary is presented in Sec. 5.
\par

\section{Multiquark operators} \lb{s2}

General gauge-invariant multiquark operators are constructed
by use of path-ordered gluon-field phase factors, also
called Wilson lines, which have the form
\be \lb{2e1}
U_{\ b}^a(C_{yx}^{})=\Big(Pe^{\displaystyle{-ig\int_{C_{yx}^{}}^{}
dz^{\mu}\ T^BA_{\ \mu}^B(z)}}\Big)_{\ b}^a\ ,
\ee
where $C_{yx}$ is an oriented curve going from $x$ to $y$,
$A_{\ \mu}^B$ are the gluon fields ($B=1,\ldots,8$), $T^B$ are
the generators of the color gauge group $SU(3)$ in the fundamental
representation, $g$ is the QCD coupling constant and $P$ represents
the path-ordering operation, meaning that the gluon fields are ordered
according to their position on the line $C_{yx}$; the integration
runs along the line $C_{yx}^{}$ from $x$ to $y$
\cite{Mandelstam:1962mi,BialynickiBirula:1963,Mandelstam:1968hz,
Nambu:1978bd}. The phase factors $U(C_{yx}^{})$ are the color
parallel transporters of the theory along the lines $C_{yx}^{}$
\cite{Corrigan:1978zg}. Examples of line $C_{yx}^{}$ are shown
in Fig. \rf{f1}; straight lines are generally chosen for their
simplicity and also for their adequacy in lattice calculations
\cite{Wilson:1974sk}.
\bfg 
\bc
\epsfig{file=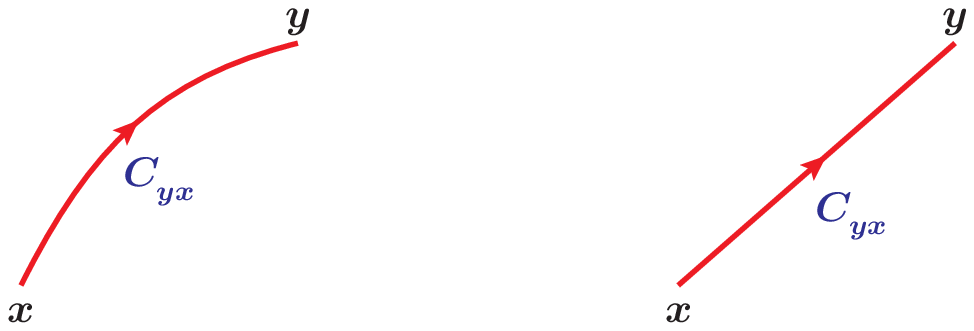,scale=0.8}
\caption{Examples of line $C_{yx}^{}$.}  
\lb{f1}
\ec
\efg
\par
For mesonic and baryonic gauge-invariant operators, one has the
following constructions:
\bea
\lb{2e2}
& &M=\overline q_a^{}(y)U_{\ b}^a(C_{yx}^{})q^b(x),\\
\lb{2e3}
& &B=\epsilon_{abc}^{}U_{\ d}^a(C_{xy}^{})q^d(y)U_{\ e}^b(C_{xt}^{})
q^e(t)U_{\ f}^c(C_{xz}^{})q^f(z),
\eea
quark flavor and spin indices being omitted and where $\epsilon$ is
the three-dimensional Levi-Civita symbol. A corresponding
pictorial representation is given in Fig. \rf{f2}.
\par
\bfg 
\bc
\epsfig{file=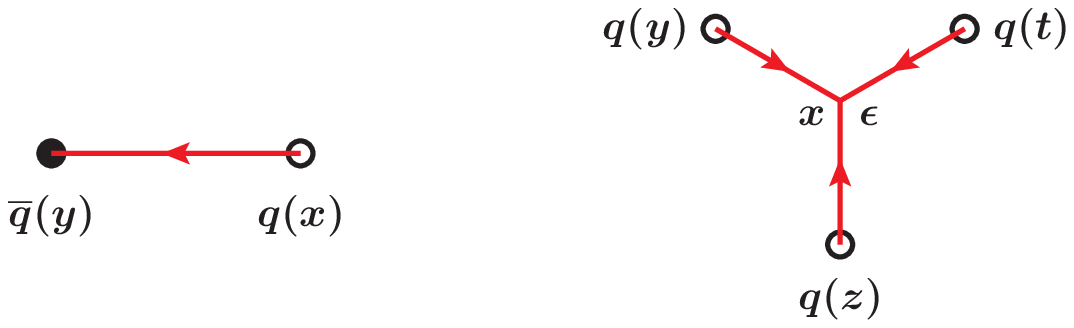,scale=0.9}
\caption{Pictorial representation of mesonic and baryonic
gauge-invariant operators, with phase factors running along straight
lines; in the baryonic case, $\epsilon$ is the Levi-Civita symbol,
put here as a reminder of the completely antisymmetric structure of
the junction vertex of the three phase-factor lines.}
\lb{f2}
\ec
\efg
\par
The choice of the line types in representations (\rf{2e2}) and
(\rf{2e3}) is arbitrary, provided they remain continuous with
rather smooth variations. Physical quantities should not depend
on that choice, which would show up only in the corresponding
wave function of states. This can be verified in the case of
bound-state energies, which are controlled by the properties of
Wilson loops at large time separations \cite{Wilson:1974sk}.
In QCD, Wilson-loop averages are expected to respect the area
law for large rectangular contours and the minimal surface property
for more general contours \cite{Wilson:1974sk,Makeenko:1980wr,
Jugeau:2003df}. Deformations of the phase-factor lines inside
the states are completely projected, when the time interval goes
to infinity, onto the wave functions (cf. the end of Appendix A
of Ref. \cite{Jugeau:2003df}).
\par
Similar constructions can be done for the multiquark operators.
Pictorial representations of tetraquark, pentaquark, and
hexaquark operators are shown in Fig. \rf{f3}. (For simplicity,
the case of hexaquark operators with three quark and three
antiquark fields will be omitted in this article; it can be
treated in a similar way as the other cases considered here.)
\bfg
\bc
\epsfig{file=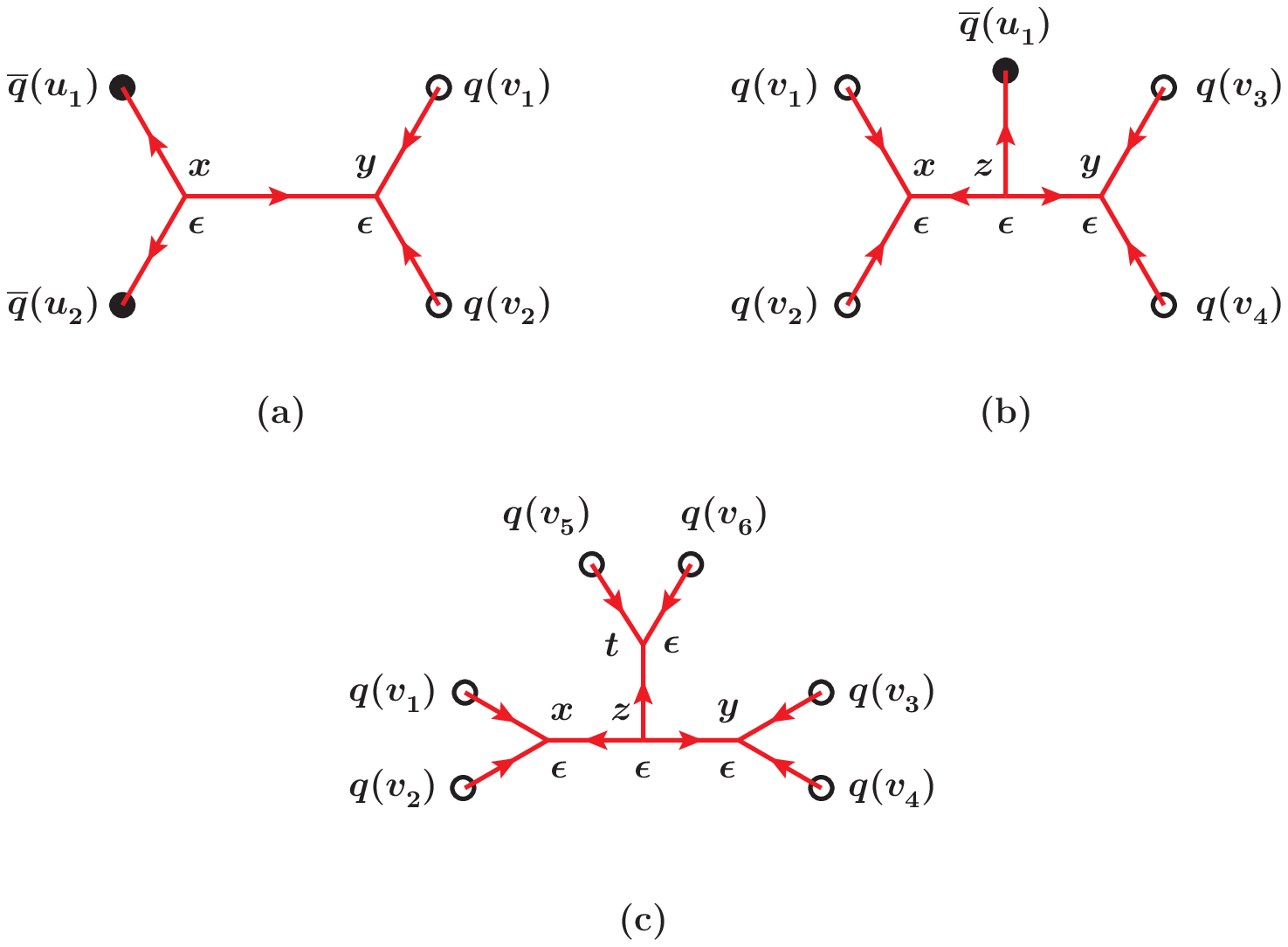,scale=0.8}
\caption{Pictorial representation of (a) tetraquark, (b) pentaquark,
and (c) hexaquark operators. Hexaquark operators with three quark
and three antiquark fields are omitted.}
\lb{f3}  
\ec
\efg
\par
The proof of the cluster reducibility property of multiquark
operators is based on the observation that the path-ordered
phase factors are elements of the color gauge group $SU(3)$.
Therefore, they satisfy the group product law
\be \lb{2e4}
U_{\ b}^a(C_{zy}^{})U_{\ c}^b(C_{yx}^{})=U_{\ c}^a(C_{zyx}^{}),
\ee
with the determinant of their matrix representation equal to 1
(cf. Appendix C of \cite{Migdal:1984gj}):
\be \lb{2e5}
\mathrm{det}(U(C_{yx}^{}))=1
=\frac{1}{3!}\ \epsilon_{a_1^{}a_2^{}a_3^{}}^{}
\epsilon^{b_1^{}b_2^{}b_3^{}}U_{\ b_1^{}}^{a_1^{}}(C_{yx}^{})
U_{\ b_2^{}}^{a_2^{}}(C_{yx}^{})U_{\ b_3^{}}^{a_3^{}}(C_{yx}^{}).
\ee
\par
A pictorial representation of Eq. (\rf{2e4}) is shown in Fig.
\rf{f4}.
\bfg 
\bc
\epsfig{file=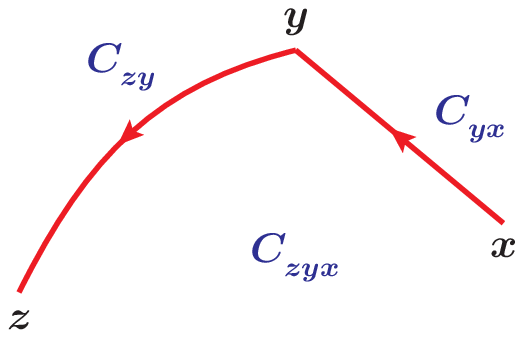,scale=0.9}
\caption{Group product law of phase factors: the product of the
two phase factors along the lines $C_{yx}^{}$ and $C_{zy}^{}$,
respectively, is equal to the phase factor along the composite line
$C_{zyx}^{}$.}
\lb{f4}
\ec
\efg
\par
We consider here the example of the tetraquark operator
(Fig. \rf{f3}(a)), which has the expression
\bea \lb{2e6}
T&=&\overline q_{a_1^{}}(u_1^{})U_{\ c_1^{}}^{a_1^{}}(C_{u_1^{}x})
\overline q_{a_2^{}}(u_2^{})U_{\ c_2^{}}^{a_2^{}}(C_{u_2^{}x})
\nonumber\\
& &\times\epsilon^{c_1^{}c_2^{}c_3^{}}
U_{\ c_3^{}}^{d_3^{}}(C_{yx}^{})\epsilon_{d_1^{}d_2^{}d_3^{}}^{}
U_{\ b_2^{}}^{d_2^{}}(C_{yv_2^{}})q^{b_2^{}}(v_2^{})
U_{\ b_1^{}}^{d_1^{}}(C_{yv_1^{}})q^{b_1^{}}(v_1^{}).
\eea
One multiplies $T$ with the determinant (\rf{2e5}),
corresponding to a phase factor taken along a line $\Gamma_{xy}^{}$,
joining $y$ to $x$. The shape of $\Gamma$ is arbitrary, but for
simplifying purposes, it could be taken of the same form as
the line $C_{yx}^{}$, with opposite orientation. Upon
using identities of the type
\be \lb{2e7}
\epsilon^{f_1^{}f_2^{}f_3^{}}\epsilon_{a_1^{}a_2^{}a_3^{}}^{}
=\delta_{\ a_1^{}}^{f_1^{}}\delta_{\ a_2^{}}^{f_2^{}}
\delta_{\ a_3^{}}^{f_3^{}}+\sum_{k_i^{}}(-1)^p
\delta_{\ a_1^{}}^{f_{k_1^{}}^{}}\delta_{\ a_2^{}}^{f_{k_2^{}}^{}}
\delta_{\ a_3^{}}^{f_{k_3^{}}^{}},
\ee
where the sum runs over all permutations of the indices $k_i$
with $(-1)^p$ representing the parity of the permutations, and
choosing each time one $\epsilon$ from the $T$ and another from
the determinant, one ends up with the expression
\newpage
\bea \lb{2e8}
T&=&\ \ \mathrm{tr}\Big(U(\Gamma_{xy}^{})U(C_{yx}^{})\Big)
\Big[\overline q(u_1^{})U(C_{u_1^{}x})U(\Gamma_{xy}^{})
U(C_{yv_1^{}})q(v_1^{})\Big]\nonumber\\
& &\ \ \ \ \ \ \times
\Big[\overline q(u_2^{})U(C_{u_2^{}x})U(\Gamma_{xy}^{})
U(C_{yv_2^{}})q(v_2^{})\Big]\nonumber \\
& &\ \ \ -\Big[\overline q(u_1^{})U(C_{u_1^{}x})U(\Gamma_{xy}^{})
U(C_{yx}^{})U(\Gamma_{xy}^{})U(C_{yv_1^{}})q(v_1^{})\Big]
\nonumber\\
& &\ \ \ \ \ \ \times
\Big[\overline q(u_2^{})U(C_{u_2^{}x})U(\Gamma_{xy}^{})
U(C_{yv_2^{}})q(v_2^{})\Big]\nonumber \\
& &\ \ \ -\Big[\overline q(u_1^{})U(C_{u_1^{}x})U(\Gamma_{xy}^{})
U(C_{yv_1^{}})q(v_1^{})\Big]\nonumber\\
& &\ \ \ \ \ \ \times
\Big[\overline q(u_2^{})U(C_{u_2^{}x})U(\Gamma_{xy}^{})
U(C_{yx}^{})U(\Gamma_{xy}^{})U(C_{yv_2^{}})q(v_2^{})\Big]
\nonumber\\
& &\ \ \ +\mathrm{tr}\Big(U(\Gamma_{xy}^{})U(C_{yx}^{})\Big)
\Big[\overline q(u_1^{})U(C_{u_1^{}x})U(\Gamma_{xy}^{})
U(C_{yv_2^{}})q(v_2^{})\Big]\nonumber\\
& &\ \ \ \ \ \ \times
\Big[\overline q(u_2^{})U(C_{u_2^{}x})U(\Gamma_{xy}^{})
U(C_{yv_1^{}})q(v_1^{})\Big]\nonumber \\
& &\ \ \ -\Big[\overline q(u_1^{})U(C_{u_1^{}x})U(\Gamma_{xy}^{})
U(C_{yv_2^{}})q(v_2^{})\Big]\nonumber\\
& &\ \ \ \ \ \ \times
\Big[\overline q(u_2^{})U(C_{u_2^{}x})U(\Gamma_{xy}^{})
U(C_{yx}^{})U(\Gamma_{xy}^{})U(C_{yv_1^{}})q(v_1^{})\Big]
\nonumber \\
& &\ \ \ -\Big[\overline q(u_1^{})U(C_{u_1^{}x})U(\Gamma_{xy}^{})
U(C_{yx}^{})U(\Gamma_{xy}^{})U(C_{yv_2^{}})q(v_2^{})\Big]
\nonumber\\
& &\ \ \ \ \ \ \times
\Big[\overline q(u_2^{})U(C_{u_2^{}x})U(\Gamma_{xy}^{})
U(C_{yv_1^{}})q(v_1^{})\Big].
\eea
The tetraquark operator is thus reexpressed in the form of a
combination of six terms, each of which is a product of 
mesonic clusters. The term with the trace operation, which appears
twice, represents a Wilson loop along the closed contour
$\Gamma_{xy}^{}C_{yx}^{}$, which is an independent
gauge-invariant operator. In case the line $\Gamma_{xy}^{}$ is
chosen of the same shape as $C_{yx}^{}$, using the generally
admitted backtracking relation \cite{Makeenko:1999hq}
\be \lb{2e9}
U_{\ b}^a(C_{yx}^{})U_{\ c}^b(C_{xy}^{})=\delta_{\ c}^a,
\ee
the Wilson loop reduces to its normalization constant (which here
assumes the value 3). Equation (\rf{2e8}) is displayed in pictorial
form in Fig. \rf{f5}, where the line $\Gamma_{xy}^{}$ has been taken
of the same shape as the line $C_{yx}^{}$ (a straight line). It is
also worthwile to notice that, if the various internal lines of the
mesonic operators are approximated by a common line of the same shape,
the decomposition will contain only two terms, corresponding to the
two different possibilities of producing the mesonic clusters.
\par
The above procedure can also be applied to the pentaquark and
hexaquark operators. For the pentaquark case, one may multiply
the operator with the determinant of the phase factor considered
along the line $yz$ of Fig. \rf{f3}(b). One obtains decompositions
into products of mesonic and baryonic operators of the type of Fig.
\rf{f6}, with other terms similar to those in Fig. \rf{f5}.
\par
\bfg 
\bc
\epsfig{file=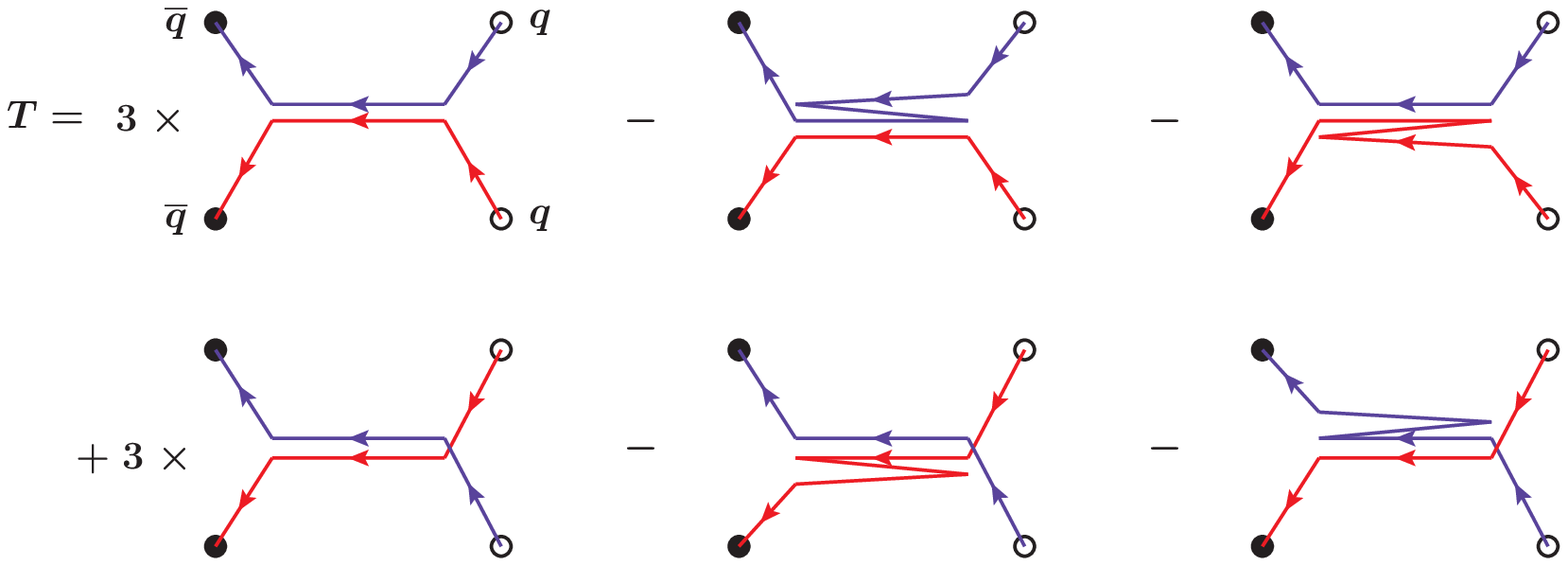,scale=0.8}
\caption{Decomposition of the tetraquark operator into a
combination of products of mesonic operators.}
\lb{f5}
\ec
\efg
\begin{figure}[h]
\bc
\epsfig{file=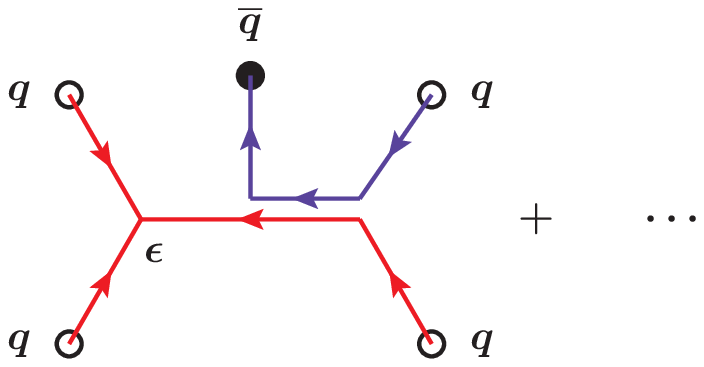,scale=0.8}
\caption{Decomposition of the pentaquark operator into a
combination of products of mesonic and baryonic operators;
the ellipsis indicates other types of the link, as in Fig. \rf{f5}.}
\lb{f6}
\ec
\end{figure}
\par
For the hexaquark case, one may multiply the operator with the
determinant of the phase factor considered along the line $yz$
of Fig. \rf{f3}(c). One obtains decompositions into products
of baryonic operators of the type of Fig. \rf{f7}, with other
terms similar to those in Fig. \rf{f5}.
\par
\bfg
\bc
\epsfig{file=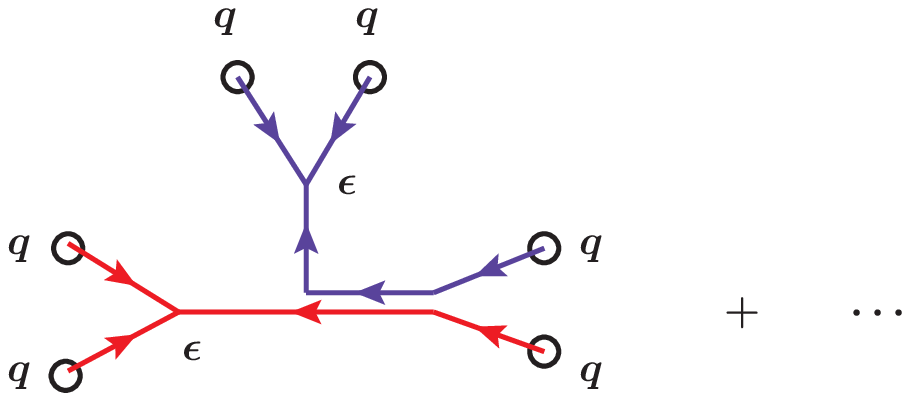,scale=0.7}
\caption{Decomposition of the hexaquark operator into a
combination of products of baryonic operators;
the ellipsis designates other types of the link, as in Fig. \rf{f5}.}
\lb{f7}
\ec
\efg
\par
As a general rule, the procedure of the insertion of the determinant
along certain lines leads to cluster reducibility whenever there
are two $Y$-shaped junctions linked by a phase-factor line. In this
connection, one observes that the usual hybrid operators, which are
obtained by insertions inside conventional mesonic and baryonic
operators of the gluon field strength $G_{\mu\nu}^BT^B$, are not
cluster reducible, since they do not contain two $Y$-shaped junctions.
Typical hybrid operators are
\bea
\lb{2e10}
\lefteqn{\hspace{-0.5 cm}
M_{hb,\mu\nu}^{}=\overline q_a^{}(y)U_{\ b}^a(C_{yz}^{})
\Big(G_{\mu\nu}^B(z)T^B\Big)_{\ c}^bU_{\ d}^c(C_{zx}^{})q^d(x),}\\
\lb{2e11}
& &{\hspace{-1.2 cm}
B_{hb,\mu\nu}^{}=\epsilon_{abc}^{}U_{\ d}^a(C_{xy}^{})q^d(y)
U_{\ e}^b(C_{xz}^{})q^e(z)U_{\ f}^c(C_{xu}^{})
\Big(G_{\mu\nu}^B(u)T^B\Big)_{\ g}^fU_{\ h}^g(C_{ut}^{})q^h(t).}
\eea
They are pictorially represented in Fig. \rf{f8}. 
\bfg 
\bc
\epsfig{file=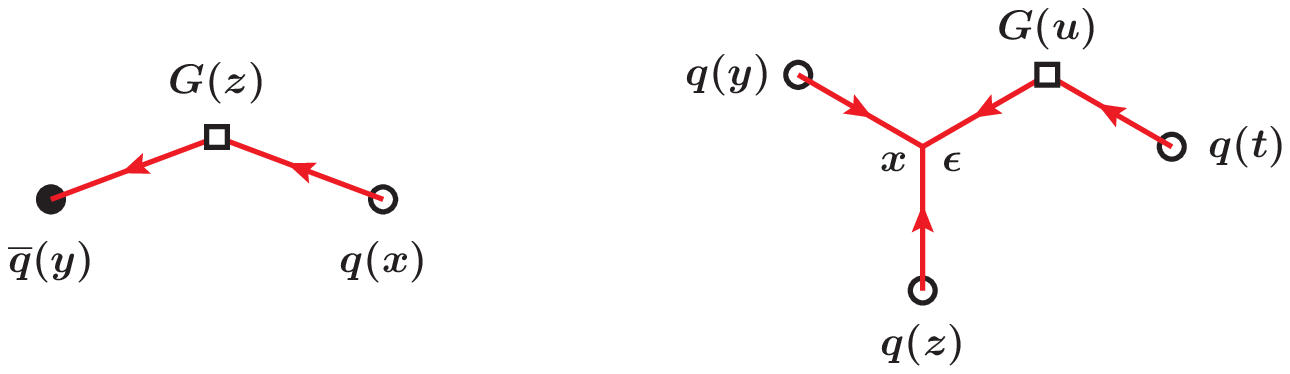,scale=0.9}
\caption{Mesonic and baryonic hybrid operators; $G$ is the gluon
field strength.}
\lb{f8}
\ec
\efg
\par

\section{{\boldmath $SU(N_c^{})$}} \lb{s3}

The color gauge group $SU(3)$ is often extended to the group
$SU(N_c^{})$, where $N_c^{}$ is the dimension of the defining
fundamental representation and is treated as a free parameter.
It turns out that, for large values of $N_c^{}$, the properties
of the theory become simplified and the theory can be studied
through an expansion in powers of the parameter $1/N_c^{}$
\cite{Coleman:1985rnk,'tHooft:1973jz,Witten:1979kh}; in particular,
inelasticity and screening effects become nondominant and the
leading terms display more clearly the confining properties of
the theory. It is therefore useful to also have representations
of multiquark operators in the $SU(N_c)$ case.
\par
While at the level of ordinary mesonic operators no changes
occur, baryonic operators undergo modifications due the
necessity of constructing completely color-antisym\-met\-ric
representations. The Levi-Civita symbol is now replaced by its
$N_c^{}$-dimensional version and this requires the junction
of $N_c^{}$ phase-factor lines. A pictorial representation of
the two operators is shown in Fig. \rf{f9}.
\bfg 
\bc
\epsfig{file=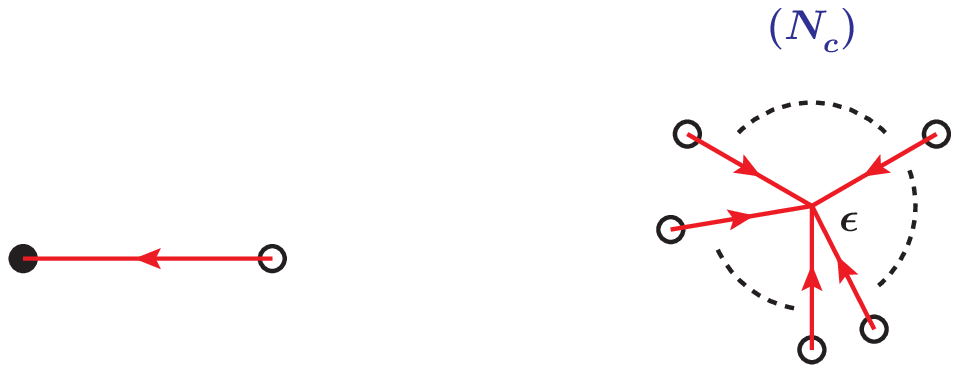,scale=1.}
\caption{Mesonic and baryonic operators in the $SU(N_c^{})$ case;
$\epsilon$ is the Levi-Civita symbol in $N_c^{}$ dimensions.}
\lb{f9}
\ec
\efg
\par
For large values of $N_c^{}$, mesons become free noninteracting
particles, with masses that are independent of $N_c^{}$ at leading
order \cite{Witten:1979kh}. The masses of baryons, due to the
increasing number of associated quarks, increase like $N_c^{}$;
however, the shapes, determined by electric charge density, and
sizes of baryons remain almost unchanged \cite{Witten:1979kh}.
\par
In a similar way, the $Y$-shaped junctions of the $SU(3)$ case are
replaced by junctions of $N_c^{}$ phase-factor lines. This feature
leads, however, to multiple choices for the construction of
multiquark operators. Considering, for instance, the tetraquark case,
Fig. \rf{f3}(a), one may replace the two external quarks and their
accompanying lines by $(N_c^{}-1)$ quarks and lines and similarly
for the antiquarks, the two junction points being linked together
by a single phase-factor line. However, one may also choose 
$2(N_c^{}-2)$ external quarks and antiquarks, with their junction
points being now linked together by two phase-factor lines, and
so forth. The extreme case corresponds to a representation where
one has two pairs of external quarks and antiquarks, with junction
points linked together by $(N_c^{}-2)$ phase-factor lines.
Thus, one obtains a sequence of tetraquark-generalizing
$[2(N_c^{}-1)]$-quark, $[2(N_c^{}-2)]$-quark, $\ldots$,
$[2(N_c^{}-(N_c^{}-2))=4]$-quark operators.
Pictorial representations of the two extremal cases are shown in
Fig. \rf{f10}.
\bfg 
\bc
\epsfig{file=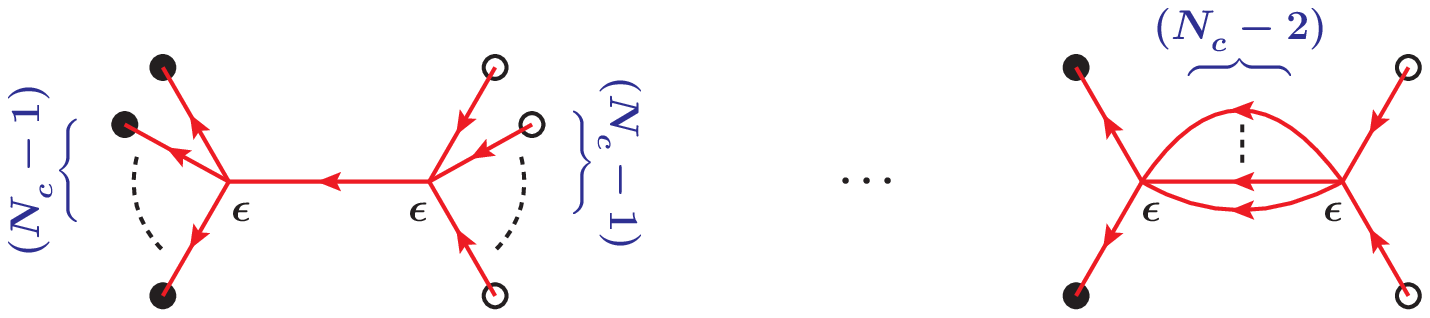,scale=1.}
\caption{Tetraquark operators in the $SU(N_c^{})$ case, where two
extreme cases are shown. The first diagram contains $(N_c^{}-1)$
quarks and $(N_c^{}-1)$ antiquarks, with a single link between
the two subsystems. The last diagram contains two quarks and two
antiquarks, with $(N_c^{}-2)$ links between the subsystems.}
\lb{f10}
\ec
\efg
\par
Each of the operators obtained above corresponds to a different
state. As is the case with ordinary baryons, it is expected that
states with increasing numbers of quarks and antiquarks will have
increasing values of masses with $N_c^{}$. The state corresponding
to the extreme case with two quarks and two antiquarks would have
the smallest mass among the many possibilities that are encountered.
\par
The pentaquark operator, like in the tetraquark case, also has
multiple extensions, as shown in Fig. \rf{f11}. In one extreme case,
one has $2(N_c^{}-1)$ external quarks and $(N_c^{}-2)$ external
antiquarks, while in the other extreme case, $(N_c^{}-1)+2$ quarks
and one antiquark. 
\bfg 
\bc
\epsfig{file=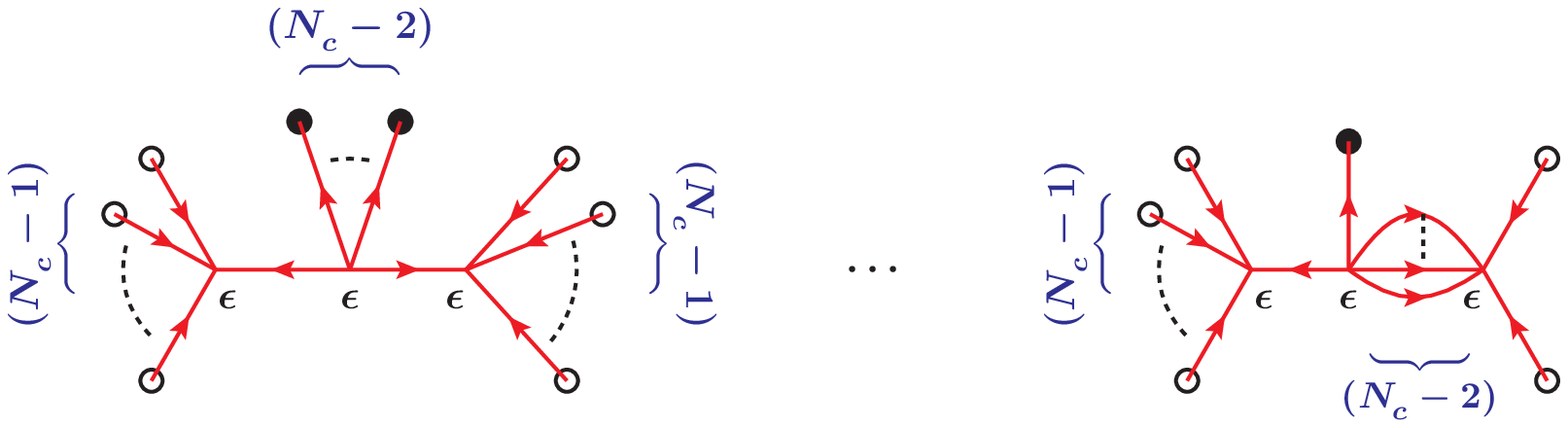,scale=0.9}
\caption{Pentaquark operators in the $SU(N_c^{})$ case, with two
extreme cases shown. The first diagram contains $2(N_c^{}-1)$
quarks and $(N_c^{}-2)$ antiquarks. The last diagram contains 
$(N_c^{}-1)+2$ quarks and one antiquark.}
\lb{f11}
\ec
\efg
\par
Hexaquark operators are constructed for general $N_c$ ($\ge 3$) as
in the previous cases. Their pictorial representations are shown in
Fig. \rf{f12}.
\bfg 
\bc
\epsfig{file=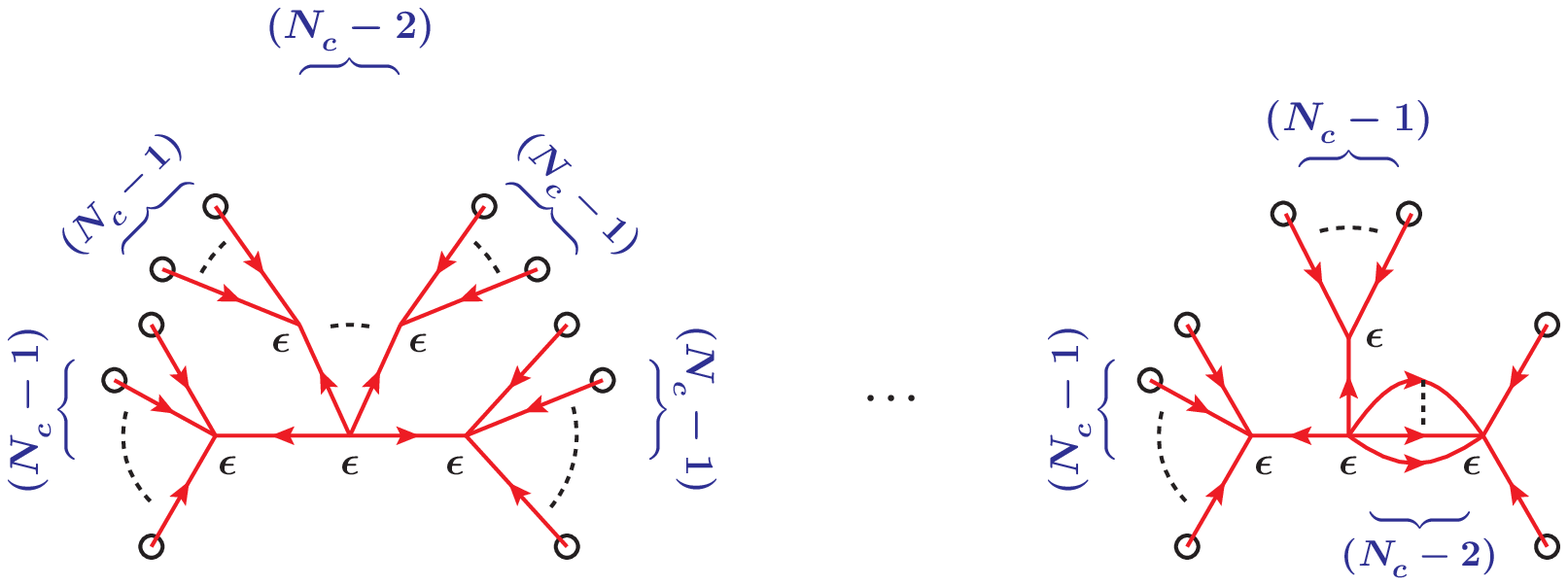,scale=0.9}
\caption{Hexaquark operators in the $SU(N_c^{})$ case, with two
extreme cases shown. The first diagram contains $N_c(N_c^{}-1)$
quarks; at the central junction point, apart from the bifurcation
into the two horizontal lines, there are bifurcations into
$(N_c^{}-2)$ lines, which in turn bifurcate each into $(N_c^{}-1)$
lines. The last diagram contains $2(N_c^{}-1)+2$ quarks.}
\lb{f12}
\ec
\efg
\par
Properties of possibly existing tetraquarks have been studied for
large values of $N_c^{}$ in Refs. \cite{Weinberg:2013cfa,
Knecht:2013yqa,Guo:2013nja,Cohen:2014tga,Maiani:2016hxw,
Lucha:2017mof,Lucha:2017gqq,Maiani:2018pef,Lucha:2018dzq}. General
cases of multiquarks have been considered in Ref.
\cite{Maiani:2018tfe}.
\par
The cluster reducibility property of multiquark operators shown
in the $SU(3)$ case can also be extended to the general $SU(N_c^{})$
case. The main ingredients of the proof, Eqs. (\rf{2e4}), (\rf{2e5})
and (\rf{2e7}), are naturally extended to that case
\cite{Migdal:1984gj}. For tetraquarks, the first operator in Fig.
\rf{f10} decomposes into a combination of products of $(N_c^{}-1)$
mesonic operators; the last operator of that figure decomposes into
a combination of products of two mesonic operators and eventually
of Wilson loops, according to the choices of the types of the
internal phase-factor lines. Similar decompositions occur for the
intermediate cases of Fig. \rf{f10}.
\par
For pentaquarks, the first operator in Fig. \rf{f11} decomposes
into a combination of products of $(N_c^{}-2)$ mesonic operators
and one baryonic operator. The last operator of that figure
decomposes into a combination of products of one mesonic operator
and one baryonic operator and eventually of Wilson loops.
\par
In the case of hexaquarks, the first operator in Fig. \rf{f12}
decomposes into a combination of products of $(N_c^{}-1)$ baryonic
operators. The last operator of that figure decomposes into a
combination of products of two baryonic operators and eventually
of Wilson loops.
\par 

\section{Discussion} \lb{s4}

Since hadronic clusters do not mutually have confining-type
interactions, one would be tempted to conclude that the global
effective interactions that govern the emergence of multiquark
bound states or resonances should be of the molecular type
\cite{Weinberg:1965zz,Voloshin:1976ap,DeRujula:1976zlg,
Amsler:2004ps,Valderrama:2012jv,Guo:2017jvc}. However, this
issue requires a more refined analysis.
\par
Molecular-type descriptions of exotic states are generally based
on effective field theories, using hadronic degrees of freedom.
These theories are essentially low-energy theories, which become less
convergent at short distances, necessitating the inclusion of
higher-order contributions with an increasing number of low-energy
constants \cite{Valderrama:2012jv,Guo:2017jvc,Weinberg:1978kz,
Gasser:1983yg,Manohar:1996cq}.
The multiquark scheme uses, in essence, quark degrees of freedom and
therefore is more adapted to describe short-distance regimes of
hadronic clusters. The main practical question that emerges is
therefore that of the determination of the domains of dominance,
inside an exotic state, of each of the preceding pictures.
\par
In order to study the internal dynamics that are at work inside
exotic states, calculations have been undertaken in the past in
lattice theory by displaying the gauge-field configurations that
are dominant in correlation functions of multiquark operators
\cite{Dosch:1982ep,Alexandrou:2004ak,Okiharu:2004wy,Okiharu:2004ve,
Suganuma:2011ci}. It turns out that, when the distance between two
hadronic clusters is smaller than the mean size of each hadron, it
is the connected $Y$-shaped-type configuration that dominates,
while in the opposite case it is the two disconnected hadronic
cluster-type configurations that are dominant. The static interquark
confining potential is then determined by the minimal value of the
total lengths of the strings of each case.
These results confirm the fact that the multiquark scheme, based
essentially on the string-junction or diquark picture, even though
globally nonconfining because of the presence or emergence of
hadronic clusters, remains a basic ingredient for a precise
description of the multiquark state. 
\par
The above results have led to the adoption of similar potential
models in nonrelativistic and semirelativistic approaches,
based on the idea of the partitioning of configuration space,
according to the dominance region of each type of potential
\cite{Lenz:1985jk,Carlson:1991zt,Martens:2006ac,Vijande:2007ix,
Vijande:2011im,Manohar:1992nd,Bicudo:2015kna,Bicudo:2016ooe,
Bicudo:2017usw,Karliner:2017qjm,Eichten:2017ffp,Quigg:2018eza}.
One of the advantages of these models is the absence of unphysical
long-range van der Waals forces, which unavoidably occur in additive
quark models with confining potentials.
\par
The concept of partitioning the configuration space, at least in
a schematic sense, according to the dominance of the various
geometrical configurations of the multiquark systems, appears as
providing the most optimal framework which is compatible with the
general cluster reducibility property of multiquark operators. 
\par
One therefore naturally arrives at the following picture of
exotic states. The latter should be described by two complementary
schemes, each valid in a separate region of configuration space:
the multiquark or diquark scheme, valid in regions where the
hadronic clusters are close to each other, and the molecular-type
scheme, valid in regions where the hadronic clusters are well
separated from each other; a crossover should prevail at
the frontier region; the weight of each configuration would
depend on the quark masses and flavors, as well as on the sectors
of quantum numbers. 
\par
According to the weight of each scheme, approximate descriptions
might be considered at a starting stage, in particular, when one
of the schemes is overdominant. In case of comparable weights,
mixtures of the two schemes offer other possibilities
\cite{Esposito:2016itg}.
\par
Recent review articles on exotic states can be found in Refs.
\cite{Guo:2017jvc,Esposito:2016noz,Lebed:2016hpi,Ali:2017jda,
Olsen:2017bmm,Karliner:2017qhf,Brambilla:2019esw}.
\par

\section{Summary} \lb{s5}

The cluster reducibility property of multiquark operators provides
a general proof of the nonexistence of completely confined or
compact multiquark states. Rather than eliminating the multiquark
scheme from the description of multiquark states, taking into
account analyses based on lattice and numerical calculations,
it streamlines the role played by the various participating
operators, according to a qualitative partitioning of configuration
space. Existing multiquark states, whether bound states or
resonances, would be schematically composed of two layers: an inner
core, having a structure governed by a connected
string-junction-type interaction, and an outer shell, having a
hadronic molecular-type structure. The weight of each layer depends
on the masses and flavors of the quarks and on the sectors of quantum
numbers that are considered. This unified scheme might provide
a better understanding of the structure of multiquark states.
\par

\section*{Acknowledgements}
This work is dedicated to the memory of Jan Stern, who has
been at the origin of the observation of the cluster reducibility
property of multiquark operators.
We thank Hans G\"unter Dosch for enlightening discussions.
D.~M.~acknowledges support from the Austrian Science Fund (FWF),
Grant No. P29028. D.~M. and H.~S. are grateful for support under
joint CNRS/RFBR Grant No. PRC Russia/19-52-15022.
The figures were drawn with the aid of the package Axodraw2
\cite{Collins:2016aya}.
\par

\end{document}